\documentclass[showpacs,amsmath,amssymb,aip,apl,reprint]{revtex4-1}

\usepackage{amsmath}    
\usepackage{graphicx}   
\usepackage{verbatim}   
\usepackage{color}      
\usepackage{subfigure}  
\usepackage{hyperref}   
\begin{document}

\title{Magnetic refrigeration with paramagnetic semiconductors at cryogenic temperatures}
\author{Alexander Vlasov}
\affiliation{McGill University, Department of Electrical and Computer Engineering, Montreal, QC, Canada, H3A 0E9}
\author{Jonathan Guillemette}
\affiliation{McGill University, Department of Physics, Montreal, QC, Canada, H3A 2T8}
\affiliation{John Abbott College, Department of Physics, Montreal, QC, Canada, H9X 3L9}
\author{Guillaume Gervais}
\affiliation{McGill University, Department of Physics, Montreal, QC, Canada, H3A 2T8}
\author{Thomas Szkopek}
\affiliation{McGill University, Department of Electrical and Computer Engineering, Montreal, QC, Canada, H3A 0E9}

\date{\today}

\begin{abstract}
We propose paramagnetic semiconductors as active media for refrigeration at cryogenic temperatures by adiabatic demagnetization. The paramagnetism of impurity dopants or structural defects can provide the entropy necessary for refrigeration at cryogenic temperatures. We present a simple model for the theoretical limitations to specific entropy and cooling power achievable by demagnetization of various semiconductor systems. Performance comparable to that of the commonly used paramagnetic salt cerous magnesium nitrate hydrate (CMN) is predicted.  
\end{abstract}

\maketitle

The adiabatic cooling of paramagnetic spins achieves refrigeration on the basis of exchange of entropy in a thermodynamic cycle. The exchange of heat by magnetization of spins was first observed in iron \cite{Warburg}, and the magnetocaloric phenomenon was first applied to magnetic refrigeration by Giauque in 1933 \cite{Giauque}. The methods and practice of magnetic refrigeration have since been developed over many years \cite{Pobell}, with the state-of-the-art involving the adiabatic demagnetization of electron spins in paramagnetic salt powders at the milliKelvin temperature scale, and nuclear spins in metals at the sub-milliKelvin temperature scale. There is renewed interest in developing new methods and materials for cryogenic refrigeration, driven by the development of new technologies such as quantum information processing \cite{Hornibrook, Mohseni} that rely upon the generation of cryogenic temperatures. 

New materials for adiabatic demagnetization are being developed, such as the intermetallic compound YbPt$_2$Sn, which combines high specific entropy at $T <$ 2~K with metallic conductivity \cite{Jang}. Fundamental extensions to refrigeration by adiabatic demagnetization are also being developed. For example, continuous cycle cooling can be achieved by the exchange of spin entropy via a spin current, leading to the spin Peltier effect. Cooling has been experimentally demonstrated in a mesoscopic device \cite{Flipse}, spurring interest in the field of spin caloritronics \cite{Bauer}. Spin entropy is believed to play a dominant role in the large thermopower of the transition metal oxide Na$_{x}$Co$_2$O$_4$ at cryogenic temperatures \cite{Wang}. The enhancement of magnetocaloric effects at quantum critical points, where changes in entropy induced by magnetic field are large, is an active area of research \cite{wolf}. Another extension to the principle of refrigeration by manipulation of entropy with magnetic field is refrigeration by adiabatic \textit{magnetization} of superconductors \cite{Giazotto}.

\begin{figure}
		\includegraphics[width=0.4\textwidth]{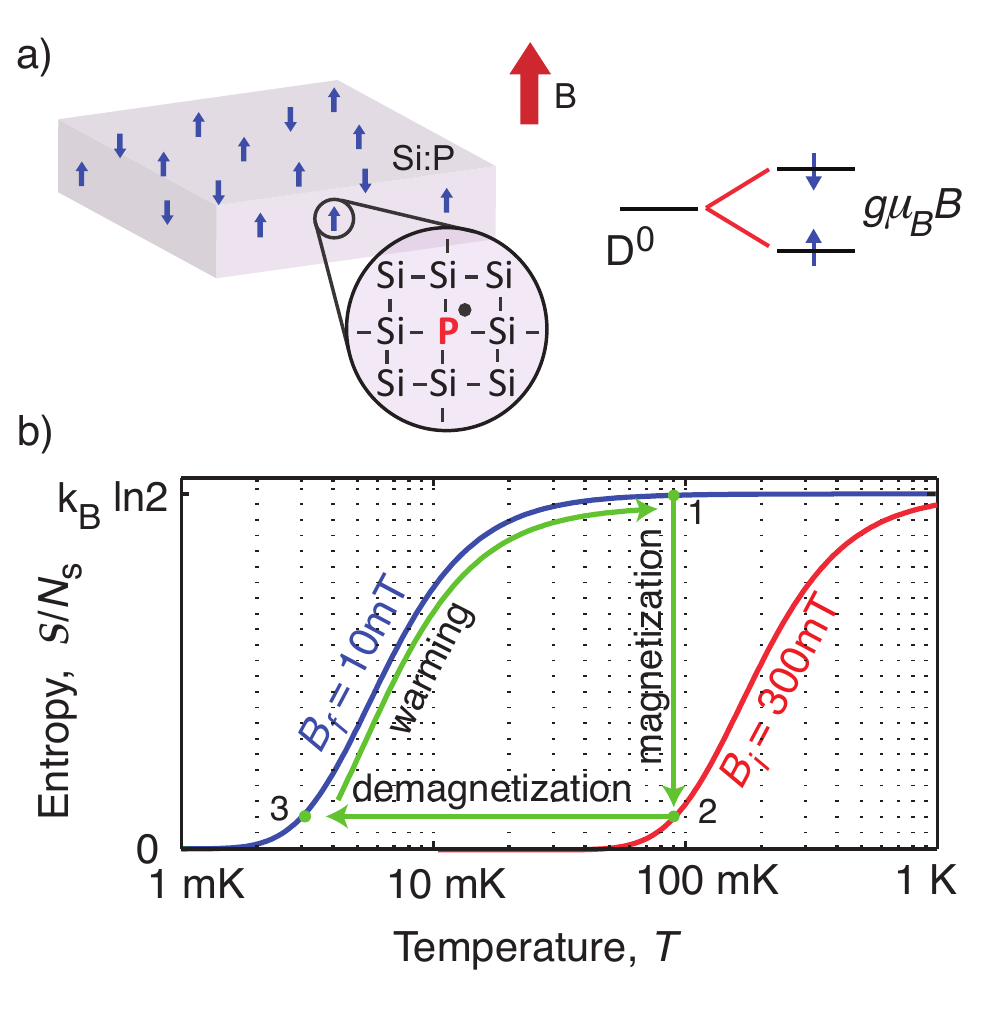} 
		\caption{ a) A doped semiconductor, such as Si:P, exhibits paramagnetism at low temperature. Un-ionized donor states $D^0$ exhibit Pauli paramagnetism, with a Zeeman splitting of $g\mu_B B$ in the case of a spin-1/2 donor atom. b) Paramagnetic entropy enables a refrigeration cycle, illustrated here with $B_i = 300~\mathrm{mT}$, $B_f = 10~\mathrm{mT}$ and a spin-1/2 system. From an initial state (1) at a temperature $T_i$, a large field $B_i$ leads to (isothermal) magnetization to a low entropy state (2). Adiabatic demagnetization to a field $B_f$ leads to a low entropy state (3) at a low temperature $T_f = T_i \cdot B_f/B_i$, cooling the paramagnetic refrigerant. Warming of the paramagnetic refrigerant returns the system to its initial state (1). }
\end{figure}

We propose paramagnetic semiconductors, such as Si:P shown schematically in Fig. 1a), as the active medium for refrigeration by adiabatic demagnetization. Paramagnetism can originate from un-ionized impurity dopants or structural defects, both of which lead to localized magnetic moments. The thermodynamic cycle is that of a typical adiabatic demagnetization refrigeration cycle, as shown in Fig.1b. The spins of impurity dopants are polarized at a high field $B_i$ at an initial temperature $T_i$ to achieve a state of low entropy $\mathcal{S}(B_i/T_i)$. An adiabatic sweep of magnetic field to a final $B_f < B_i$ maintains constant entropy $\mathcal{S}(B_f/T_f)=\mathcal{S}(B_i/T_i)$, thus bringing spins to a final temperature $T_f = T_i \cdot B_f/B_i$. The reservoir of low temperature spins can thus be used as a refrigerant. Ever present diamagnetism does not make a contribution to field dependent entropy because diamagnetic magnetization $M$ is temperature independent, and Maxwell's relation implies $(\partial \mathcal{S}/ \partial B)_T = (\partial M / \partial T)_B \sim 0$. Doped semiconductors have several properties that are advantageous for cryogenic refrigeration. The paramagnetic entropy $\mathcal{S}(B/T)$ is proportional to the concentration of impurity dopant spins. There is a wide variety of semiconductor hosts and impurity dopants, giving a wide range of experimentally accessible impurity dopant properties and concentrations. Lastly, semiconductors are the most common substrate used for electronic and micro-electromechanical (MEMS) devices, and thus offers the possibility of monolithic integration of solid state refrigerant with active devices. Silicon is particularly appealing because of its wide availability and well understood properties.

We develop a simple theoretical model for refrigeration by semiconductor paramagnetism, focussing our discussion on impurity doped semiconductors although similar principles apply to semiconductors with structural defects. An essential requirement for effective cooling is that the paramagnetic specific heat of donor bound electron spins (or acceptor bound hole spins) dominates over phonon specific heat and itinerant electron specific heat. This condition has been experimentally shown to be satisfied at low temperatures and moderate magnetic fields in heavily doped silicon, for example \cite{kobayashi, lakner94, wagner97}. The phonon specific heat is $C_{phonon} =  (12/5)\cdot ( \pi^4 k_B N_A/V_m) \cdot (T/\Theta_D)^3$, where $N_A$ is Avogadro's number, $\Theta_D$ is the Debye temperature of the semiconductor, $V_m$ is the molar volume and $k_B$ the Boltzmann constant. The paramagnetic specific heat is taken to be that of an ideal ensemble of non-interacting spin-1/2 impurity dopants, leading to the Schottky form \cite{pathria}, 
\begin{equation}
C_{spin} = N_{s} k_B \cdot \left( \frac{ g \mu_B B }{ 2 k_B T } \right)^2 \cdot  \mathrm{sech}^2 \left( \frac{g \mu_B B }{ 2 k_B T } \right),
\label{Cspin}
\end{equation}
where $N_{s}$ is the spin density, $\mu_B$ is the Bohr magneton and $g$ the Land\'e g-factor. The generalization to spin-$J$ impurity dopants is straightforward \cite{Pobell}. Itinerant electrons in semiconductors with impurity dopant concentrations exceeding the critical density of the metal-insulator transition (MIT) leads to a third contribution to specific heat, $C_{electron} = \pi^{2/3} \cdot k_B^2 T \cdot DOS(E_F)$, where $DOS(E_F)$ is the density of states (DOS) at the Fermi energy.

\begin{figure}
		\includegraphics[width=0.4\textwidth]{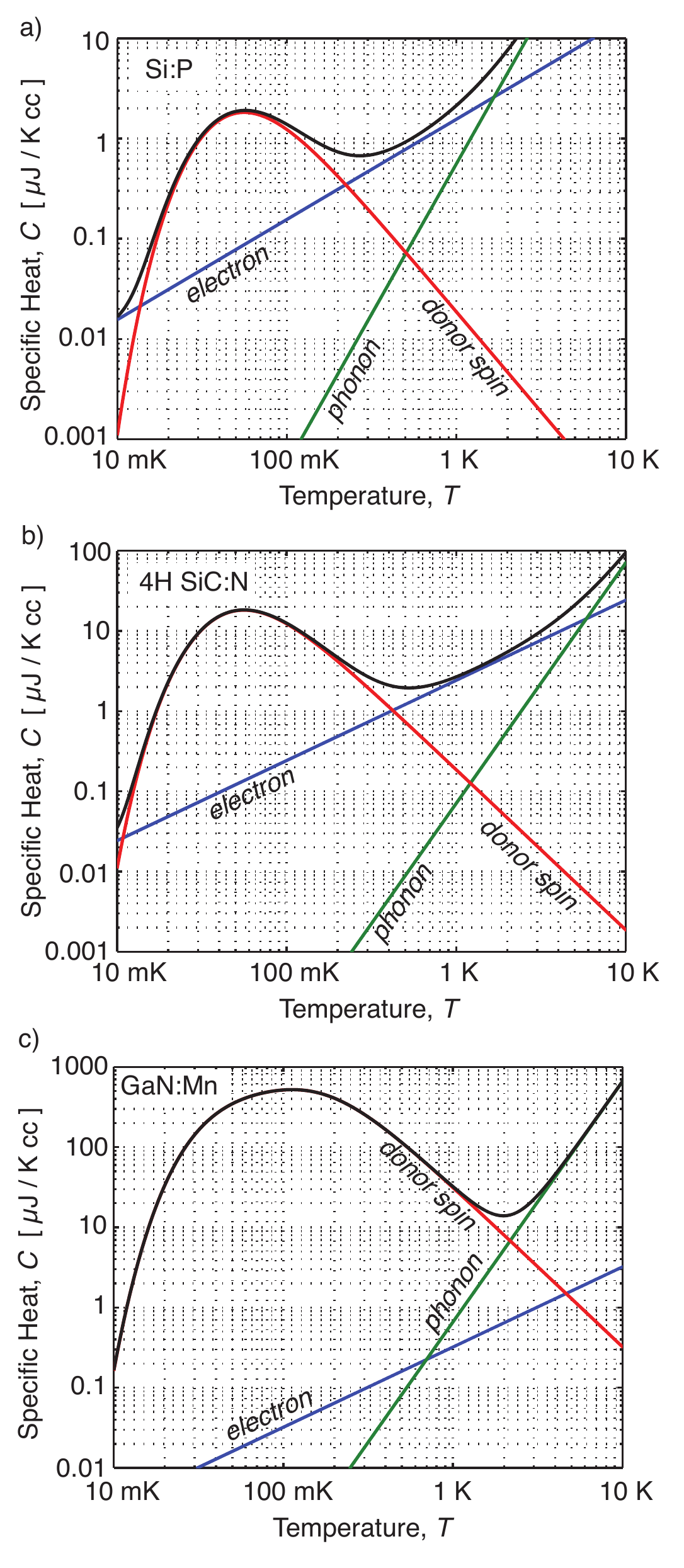} 
		\caption{ The theoretical specific heat $C$ of several heavily doped semiconductor systems, including contributions from paramagnetism (red line), itinerant electrons (blue line) and phonons (green line). The applied magnetic field is $B = 100~\mathrm{mT}$ and the spin density $N_s$ is taken from experimental measurements. a) Si:P at a P density $N_D=1.0\times10^{18}\mathrm{cm}^{-3}$ and corresponding spin-1/2 density $N_s = 3\times10^{17}\mathrm{cm}^{-3}$ as measured by specific heat \cite{lakner94}. The free electron density was estimated as $n = N_D - N_s$. b) 4H-SiC:N at an N density $N_D=7.5\times10^{18}\mathrm{cm}^{-3}$ and corresponding spin-1/2 density $N_s = 3.0\times10^{18}\mathrm{cm}^{-3}$ as measured by electron spin resonance \cite{zabrodskii}. The free electron density was estimated as $n = N_D - N_s$. c) GaN:Mn at an Mn density $N_D=4.45\times10^{19}\mathrm{cm}^{-3}$ and corresponding spin-5/2 density $N_s = N_D$ as measured by magnetization \cite{zajac}. The free electron density was measured to be $n \leqslant 1.0\times10^{18}\mathrm{cm}^{-3}$.
		 }
\end{figure}

\begin{figure*}[t]
		\includegraphics[width=1.0\textwidth]{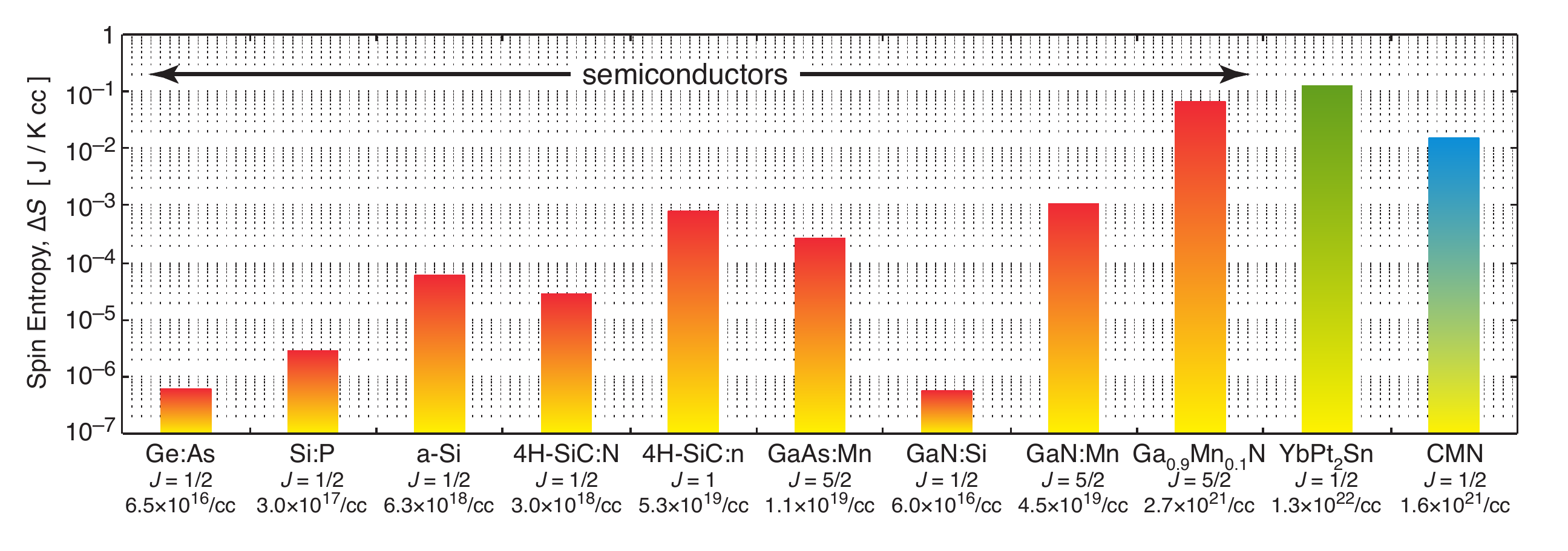} 
		\caption{ The theoretical spin entropy $\Delta \mathcal{S}= k_B N_s \ln ( 2J + 1 )$ of several semiconductor systems is compared along with the intermetallic compound YbPt$_2$Sn \cite{Jang} and the commonly used paramagnetic salt CMN \cite{giauqueCMN}. The experimentally measured spin-$J$ and spin density $N_s$ are indicated for each material system. The semiconductor systems shown include Ge:As \cite{zabrodskii}, Si:P \cite{lakner94}, a-Si \cite{stutzmann}, 4H-SiC:N \cite{zabrodskii}, neutron irradiated 4H-SiC \cite{wang}, GaAs:Mn \cite{pawlowski}, GaN:Si \cite{wolos}, GaN:Mn \cite{zajac} and Ga$_{1-x}$Mn$_x$N \cite{zajac}.
		 }
\end{figure*}

There are several heavily doped semiconductors in which paramagnetism has been observed. As will be discussed further below, the density $N_s$ of impurity bound spins depends upon the impurity dopant concentration $N_D$ in a non-trivial manner. Theoretical specific heat curves at a modest magnetic field of $B = 100~\mathrm{mT}$ are shown in Fig. 2 based on \textit{experimentally} measured spin densities $N_s$ for a variety of semiconductor systems. Fig. 2a) shows the theoretical specific heat of Si:P with the maximum spin-1/2 density reported to date, $N_s = 3\times10^{17}\mathrm{cm}^{-3}$ at a P doping density $N_D=1\times10^{18}\mathrm{cm}^{-3}$, as inferred from specific heat measurements \cite{lakner94}. Fig. 2b) shows the theoretical specific heat of 4H-SiC:N with the maximum spin-1/2 density reported to date, $N_s = 3\times10^{18}\mathrm{cm}^{-3}$ at an N doping density $N_D=7.5\times10^{18}\mathrm{cm}^{-3}$, as measured by electron spin resonance \cite{zabrodskii}. Fig. 2c) shows the theoretical specific heat of GaN:Mn at an Mn density $N_D=4.45\times10^{19}\mathrm{cm}^{-3}$, where the spin-5/2 density $N_s = N_D$ follows ideal paramagnetic behaviour as seen in direct measurement of magnetization \cite{zajac}. The paramagnetic specific heat dominates over both phonon and electron contributions at temperatures of $T\sim 1~\mathrm{K}$, and thus magnetic field can be used to manipulate the entropy of these systems at $T \leqslant 1~\mathrm{K}$.

To elucidate paramagnetic refrigeration performance, it is useful to consider the explicit form of the paramagnetic entropy, again for non-interacting spin-1/2 impurity dopants \cite{pathria},
\begin{equation}
\label{eq:S}
\mathcal{S} =  N_s k_{B} \left[ \ln \left \{ 2 \cosh \left( x \right) \right \} -  x \tanh \left( x \right) \right]
\end{equation}
where the dimensionless ratio $x =g \mu_B B / 2 k_B T$ determines the paramagnetic entropy per spin $\mathcal{S}/N_s$. The general case of a spin-$J$ impurity is more complex, where $J$ and $x$ determine $\mathcal{S}/N_s$ \cite{Pobell}. The most commonly used impurity dopants give spin-1/2 paramagnetism. There is a wide range of electron g-factors accessible in semiconductors, from $g = +2$ in silicon \cite{feher} through to $g < -10$ in III-V compound semiconductors with large spin-orbit coupling \cite{kosaka}. Because the cooling cycle occurs under ideal adiabatic conditions where $\Delta \mathcal{S} = 0$, the practical implication of g-factor selection is that it determines the magnetic field $B_i$ that is required to polarize spins into a low entropy state at the initial temperature $T_i$.

The spin density $N_s$ is the most critical parameter that determines the cooling characteristics of the refrigerator. The maximum paramagnetic spin-$J$ entropy that is available for refrigeration is $\Delta \mathcal{S}=N_s k_B \ln ( 2 J +1)$. The maximum heat $\Delta Q$ that can be absorbed by the refrigerant at the final field $B_f$ is,
\begin{equation}
\Delta Q = \int_{T_f}^{\infty} C_{spin} dT =  J N_s |g|\mu_B B_f
\end{equation}
in the limit of a highly polarized state of the refrigerant, $k_B T_f \ll |g|\mu_B B_f$, prior to warming. In other words, the maximum heat that can be absorbed by $N_s$ spins is the energy required to fully depolarize the $N_s$ spins at the terminal field $B_f$.

Importantly, electron spin relaxation times $T_1$ in semiconductors can vary by many orders of magnitude, decreasing to short times in the case of heavily doped semiconductors. For example, $T_1 < 1~\mathrm{s}$  for $N_s > 10^{17}~\mathrm{cm}^{-3}$ in silicon \cite{fehergere}. Over the time scale of a demagnetization, typically of the order of $10^2$~s to $10^4$~s, the impurity spins and lattice phonons are effectively in thermal equilibrium, and a single temperature $T$ describes the semiconductor lattice and spins. The rapid thermal equilibration time admits the possibility of \textit{isothermal} heat extraction by demagnetization of the semiconductor. In the isothermal limit, the rate of heat extraction $\dot Q$ from the environment by the semiconductor refrigerant is $\dot Q = T\dot S = T \left( \partial S / \partial B \right)_{T} \dot B$. In the case of the ideal spin-1/2 paramagnetic system, the heat extraction rate reaches a peak value of $\dot Q_{peak} = 0.448 \cdot N_s \mu \dot B$ at a magnetic field $\mu_B B_{peak} = 0.772 \cdot k_{B}T$.

The density $N_s$ of impurity bound spins, which is critical to refrigerator performance, itself depends upon the impurity dopant concentration $N_D$.  At impurity dopant concentrations well below the critical density $N_c$ of the MIT, single electrons are bound to impurity dopants and $N_s = N_D$. As the impurity concentration approaches the critical density $N_c$ , the spin density $N_s$ exhibits a maximum before decreasing with increasing $N_D$. Experimental measurements of specific heat in Si:P \cite{lakner94}, electron spin resonance in Ge:As and 4H-SiC:N \cite{zabrodskii}, and electron spin resonance in GaN:Si \cite{wolos} all show a rapid drop in spin density $N_s$ as the dopant density approaches the critical density $N_c$. The microscopic mechanisms responsible for the loss of paramagnetism as $N_D$ approaches $N_c$ varies from one material system to another. Anti-ferromagnetic ordering has been postulated in the form of singlet formation at doubly occupied donor sites \cite{wolos} and singlet formation of spatially proximate donor sites \cite{zabrodskii, lakner94}, per the theory of Bhatt and Lee \cite{bhatt}. From a thermodynamic perspective, the onset of spin ordering depletes the spin entropy that would otherwise be available for use in a magnetic refrigeration scheme. At doping densities above $N_c$, electron delocalization reduces the density of localized electrons that contribute directly to paramagnetism. The Pauli paramagnetism of the delocalized electron gas above the MIT gives a temperature independent magnetization, and thus provides no entropy for magnetic refrigeration per the relation $(\partial \mathcal{S}/ \partial B)_T = (\partial M / \partial T)_B \sim 0$. 

A summary of the specific spin entropy $\Delta \mathcal{S} = N_s k_{B} \ln (2J+1)$ available for magnetic refrigeration for a variety of material systems is shown in Fig. 3, on the basis of the \textit{experimentally measured} spin density $N_s$ and angular momentum $J$. Several common doped semiconductor systems are compared, including Ge:As \cite{zabrodskii}, Si:P \cite{lakner94}, 4H-SiC:N \cite{zabrodskii}, GaAs:Mn \cite{pawlowski}, GaN:Si \cite{wolos} and GaN:Mn \cite{zajac}. Among these doped semiconductors, the general trend is that a greater spin density is achievable in larger gap semiconductors. The critical dopant density $N_c$ at which electron delocalization occurs is very well approximated by $N_c = (0.26/a_B^*)^3$ for a wide range of semiconductors \cite{edwards}, where $a_B^*$ is an effective Bohr radius that tends to smaller values for larger gap semiconductors and larger impurity ionization energies. In short, tighter confinement of localized states enables a greater density of localized states without the onset of delocalization.

Paramagnetism from localized electronic states can be introduced into semiconductors with structural defects. Amorphous silicon is typically prepared in a hydrogenated state, a-Si:H, to passivate dangling bonds. The residual spin density is $N_s \sim 10^{16}\mathrm{cm}^{-3}$, but can be significantly enhanced to a density $N_s = 6.3\times10^{18}\mathrm{cm}^{-3}$ by thermal annealing to remove passivating hydrogen atoms and produce a-Si \cite{stutzmann}. Neutron irradiation offers another alternative method of imparting paramagnetism to semiconductors \cite{wang}, with a spin-1 density of $N_s = 5.3\times10^{19}\mathrm{cm}^{-3}$ achieved in 4H-SiC. The paramagnetism of neutron irradiated 4H-SiC is attributed to divacancies $\mathrm{V_{Si}V_{C}}$ with a magnetic moment of $2\mu_B$ \cite{wang}. Lastly, we note that significant paramagnetism can persist in dilute magnetic semiconductors such as Ga$_{1-x}$Mn$_x$N \cite{zajac}, where spin-5/2 paramagnetism of Mn at an effective concentration $x_{\mathrm{eff}} = 0.061$ was observed in a crystal with Mn fraction $x=0.096$. Anti-ferromagnetic coupling reduces the paramagnetism of magnetic semiconductors, but nonetheless enables a spin-5/2 density of $N_s =2.7 \times 10^{21}\mathrm{cm}^{-3}$ to be achieved. For comparison, a commonly used paramagnetic salt for magnetic cryo-refrigeration is cerous magnesium nitrate hydrate (CMN)\cite{giauqueCMN}, where the Ce$^{3+}$ ions give a spin-1/2 density of $N_s = 1.6\times10^{21}~\mathrm{cm}^{-3}$. The magnetic semiconductor system Ga$_{1-x}$Mn$_x$N gives a higher density of spins with higher momentum $J$ than that of CMN.

We finally consider the minimum temperature $T_f$ that is achievable by adiabatic demagnetization. There is a critical temperature $T_c$ below which the spin system undergoes a transition to an ordered state, depleting entropy available for refrigeration and setting a limit to the achievable base temperature \cite{Pobell}. In the absence of the itinerant electrons of a metallic state, it is expected that magnetic dipole-dipole interactions set the temperature scale at which ordering occurs, $k_B T_c \sim (\mu_0/4\pi) \cdot \mu_B^2 \left < 1/ r^3 \right >$, with $\left < 1/r^3 \right > \sim N_s$. There is thus a trade-off between specific spin entropy $\Delta \mathcal{S} \propto N_{s}$ and the minimum temperature $T_c \propto N_s$, while the strength of spin-spin interaction varies from material to material. Ordering temperatures are typically determined experimentally. For example, the ordering temperature of CMN is $T_c = 2$~mK\cite{Pobell} at $N_s = 1.6\times10^{21}~\mathrm{cm}^{-3}$, and the ordering temperature of the recently discovered intermetallic compound YbPt$_2$Sn is $T_c = 250$~mK\cite{Jang} at $N_s = 1.3\times10^{22}~\mathrm{cm}^{-3}$. There is a paucity of experimental work reported on the spin ordering temperature of magnetic semiconductors. With spin densities comparable to or lower than that of CMN, semiconductor paramagnetism is expected to persist down to the $\sim$~mK temperature scale.

In conclusion, the measured spin-$J$ and spin density $N_s$ of paramagnetic semiconductors provides spin entropy comparable to that of paramagnetic salts used in adiabatic demagnetization refrigerators. Paramagnetic semiconductors are thus appealing as solid state cryo-refrigerants that can be monolithically integrated with active electronic components. Integration at this scale optimizes cooling by minimizing the volume of material to be cooled, and minimizing the number of interfaces that can contribute to parasitic thermal resistance. The chemical stability of semiconductors is also favourable in comparison with hydrated paramagnetic salts. Nonetheless, further work is required to identify the most promising semiconductor system, as their is a wide variety of impurity dopants and structural defects available to choose from. Significant experimental work is required to quantify the paramagnetic contribution to specific heat of semiconductors at cryogenic temperatures, and to determine the electron spin ordering temperatures that will set the temperature limit to magnetic refrigeration in these materials.

We thank Pascal Pochet and Shengqiang Zhou for bringing to our attention recent developments in neutron irradiated semiconductors. The authors acknowledge financial support from the Canada Research Chairs Program, the Canadian Institute for Advanced Research, and the Canadian Natural Sciences and Engineering Research Council.


\begin{thebibliography}{7}

\bibitem{Warburg} E. Warburg and L. H\"onig, Ueber die w\"arme, welche durch periodisch wechselnde magnetisirende kr\"afte im eisen erzeugt wird, Annal. der Phys. 256, 814Ð835 (1883).

\bibitem{Giauque} W. F. Giauque and D. P. MacDougall, Attainment of temperatures below $1^\circ$ absolute by demagnetization of Gd$_2$(SO$_4$)$_{3}\cdot$8H2O, Phys. Rev. \textbf{43}, 768 (1933).

\bibitem{Pobell} F. Pobell, Matter and Methods at Low Temperatures (Springer, New York, 1992).

\bibitem{Hornibrook} J. M. Hornibrook, J. I. Colless, I. D. Conway Lamb, S. J. Pauka, H. Lu, A. C. Gossard, J. D. Watson, G. C. Gardner, S. Fallahi, M. J. Manfra, and D. J. Reilly, Cryogenic Control Architecture for Large-Scale Quantum Computing, Phys. Rev. Applied \textbf{3}, 024010 (2015).

\bibitem{Mohseni} M. Mohseni, P. Read, H. Neven, S. Boixo, V. Denchev, R. Babbush, A. Fowler, V. Smelyanskiy and J. Martinis, Commercialize quantum technologies in five years, Nature \textbf{543}, 171 (2017).

\bibitem{Jang} D. Jang, T. Gruner, A. Steppke, K Mistumoto, C. Geibel and M. Bando, Large magnetocaloric effect and adiabatic demagnetization refrigeration with YbPt$_2$Sn, Nature Comms. \textbf{6}, 8680 (2015).

\bibitem{Flipse} J. Flipse, F. L. Bakker, A. Slachter, F. K. Dejene and B. J. van Wees, Direct observation of the spin-dependent Peltier effect, Nature Nanotech. \textbf{7}, 166 (2012).

\bibitem{Bauer} G. E. W. Bauer, E. Saitoh and B. J. van Wees, Spin Caloritronics, Nature Mat. \textbf{11}, 391 (2012).


\bibitem{Wang} Y. Wang, N. S. Rogado, R. J. Cava and N. P. Ong, Spin entropy as the likely source of enhanced thermopower in Na$_{x}$Co$_2$O$_4$, Nature \textbf{423}, 425 (2003).

\bibitem{wolf} B. Wolf \textit{et al.}, Magnetocaloric effect and magnetic cooling near a field-induced quantum-critical point, Proc. Nat. Acad. Sci. \textbf{108}, 6862 (2011).

\bibitem{Giazotto} F. Dolcini and F. Giazotto, Adiabatic magnetization of superconductors as a high-performance cooling mechanism, Phys. Rev. B \textbf{80}, 024503 (2009).

\bibitem{kobayashi} N. Kobayashi, S. Ikehata, S. Kobayashi, and W. Sasaki, Magnetic Field Dependence of Specific Heat of Heavily Phosphorus Doped Silicon, Solid State Comm. \textbf{32}, 1147 (1979).

\bibitem{lakner94} M. Lakner, H. v. L\"ohneysen, A. Langenfeld and P. W\"olfe, Localized magnetic moments in Si:P near the metal-insulator transition, Phys. Rev. B \textbf{50}, 17064 (1994).

\bibitem{wagner97} S. Wagner, M. Lakner, and H. v. L\"ohneysen, Specifc heat of Si:(P,B) at low temperatures, Phys. Rev. B \textbf{55}, 4219 (1997).

\bibitem{pathria} R.K. Pathria, Statistical Mechanics (Butterworth-Heinemann, New York, 1996).

\bibitem{zabrodskii} A. G. Zabrodskii, Magnetic ordering in doped semiconductors near the metalÐinsulator transition, phys. stat. sol. (b) \textbf{241}, 33 (2004).

\bibitem{zajac} M. Zaj\c{a}c, J. Gosk, M. Kami\'{n}ska, A. Twardowski, T. Szyszko and S. Podsiadlo, Paramagnetism and antiferromagnetic dÐd coupling in GaMnN magnetic semiconductor, Appl. Phys. Lett. \textbf{79}, 2432 (2001).

\bibitem{feher} G. Feher, Electron Spin Resonance Experiments on Donors in Silicon. I. Electronic Structure of Donors by the Electron Nuclear Double Resonance Technique, Phys. Rev. \textbf{114}, 1219 (1959).

\bibitem{kosaka} H. Kosaka, A. A. Kiselev, F. A. Baron, K. W. Kim and E. Yablonovitch, Electron g-factor Engineering in III-V Semiconductors for Quantum Communications, Elect. Lett. \textbf{37}, 464 (2001).

\bibitem{fehergere} G. Feher and E. A. Gere, Electron Spin Resonance Experiments on Donors in Silicon. II. Electron Spin Relaxation Effects, Phys. Rev. \textbf{114}, 1245 (1959)

\bibitem{wolos} A. Wolos, Z. Wilamowski, M. Piersa, W. Strupinski, B. Lucznik, I. Grzegory, and S. Porowski, Properties of metal-insulator transition and electron spin relaxation in GaN:Si, Phys. Rev. B \textbf{83}, 165206 (2011).

\bibitem{bhatt} R. N. Bhatt and P. A. Lee, Scaling Studies of Highly Disordered Spin-1/2 Antiferromagnetic Systems, Phys, Rev. Lett. \textbf{48}, 344 (1984).

\bibitem{edwards} P. P. Edwards and M. J. Sienko, Universality aspects of the metal-nonmetal transition in condensed media, Phys. Rev. B \textbf{17}, 2575 (1978).

\bibitem{pawlowski} M. Pawlowski, M. Piersa, A. Wolos, M. Palczewska, G. Strzelecka, A. Hruban, J. Gosk, M. Kami\'{n}ska, and A. Twardowski, Mn Impurity in Bulk GaAs Crystals, Acta. Phys. Pol. \textbf{108}, 825 (2005).

\bibitem{stutzmann} M. Stutzmann and D. K. Biegelsen, Electron-spin-lattice relaxation in amorphous silicon and germanium, Phys. Rev. B \textbf{28}, 6256 (1983).

\bibitem{wang} Y. Wang, Y. Liu, E. Wendler, R. H\"ubner, W. Anwand, G. Wang, X. Chen, W. Tong, Z. Yang, F. Munnik, G. Bukalis, X. Chen, S. Gemming, M. Helm,
and S. Zhou, Defect-induced magnetism in SiC: Interplay between ferromagnetism and paramagnetism, Phys. Rev. B \textbf{92}, 174409 (2015).

\bibitem{giauqueCMN} W. F. Giauque, R. A. Fisher, E. W. Hornung, and G. E. Brodale, Magnetothermodynamics of Ce$_2$Mg$_3$(NO$_3$)$_{12}$$\cdot$24H$_2$O. I. Heat capacity, entropy, magnetic moment from 0.5 to 4.2$^\circ$K with fields to 90 kG along the a crystal axis. Heat capacity of Pyrex 7740 glass in fields to 90 kG, J. Chem. Phys. \textbf{58}, 2621 (1975).



\end{thebibliography}
\end{document}